\documentclass[12pt]{article}
\usepackage[dvips]{color}
\usepackage{epsfig}
\usepackage{amsmath}
\usepackage{graphicx}

\begin{document}

\title{\bf Interacting tachyon dark energy in non-flat universe}
\author{{M. R. Setare $^{a}$ \thanks{Email: rezakord@ipm.ir}\hspace{1mm}
, J. Sadeghi $^{b}$\thanks{Email: pouriya@ipm.ir}\hspace{1mm} and A. R. Amani$^{c}$ \thanks{Email: a.r.amani@iauamol.ac.ir}}\\
$^{a}${\small {\em Department of Science, Payame Noor University, Bijar, Iran }}\\
$^b${\small {\em Sciences Faculty, Department of Physics, Mazandaran
University,}}\\{\small {\em P .O .Box 47415-416, Babolsar, Iran}}\\
$^c$ {\small {\em  Department of Physics, Islamic Azad University - Ayatollah Amoli Branch,}}\\
        {\small {\em P.O.Box 678, Amol, Iran}}}
\maketitle

\begin{abstract}
In this paper we study the tachyon cosmology in non-interacting and
interacting cases in non-flat FRW universe. Then we reconstruct the
potential and the dynamics of the tachyon field which describe
tachyon cosmology.

\end{abstract}
\section{Introduction}
Nowadays it is strongly believed that the universe is experiencing
an accelerated expansion. Recent observations from type Ia
supernovae \cite{SN} in associated with Large Scale Structure
\cite{LSS} and Cosmic Microwave Background anisotropies \cite{CMB}
have provided main evidence for this cosmic acceleration. In order
to explain why the cosmic acceleration happens, many theories have
been proposed. Although theories of trying to modify Einstein
equations constitute a big part of these attempts, the mainstream
explanation for this problem, however, is known as theories of dark
energy. It is the most accepted idea that a mysterious dominant
component, dark energy, with negative pressure, leads to this cosmic
acceleration, though its nature and cosmological origin still remain
enigmatic at present. \\
The most obvious theoretical candidate of dark energy is the
cosmological constant $\lambda$ (or vacuum energy)
\cite{Einstein:1917,cc} which has the equation of state $\omega=-1$.
An alternative proposal for dark energy is the dynamical dark energy
scenario.  So far, a large class of scalar-field dark energy models
have been studied, including quintessence \cite{quintessence},
K-essence \cite{kessence}, tachyon \cite{tachyon}, phantom
\cite{phantom}, ghost condensate \cite{ghost1,ghost2} and quintom
\cite{quintom}, interacting dark energy models \cite{intde},
braneworld models \cite{brane}, and Chaplygin gas models \cite{cg1},
etc.
 An interacting tachyonic-dark
matter model has been studied in Ref. {\cite{c10}}.\\
In this paper, we consider the issue of the tachyon as a source of
the dark energy. The tachyon is an unstable field which has become
important in string theory through its role in the Dirac-Born-Infeld
(DBI) action which is used to describe the D-brane action \cite
{7,8}. It has been noticed that the cosmological model based on
effective lagrangian of tachyon matter
\begin{equation} \label{lag}
L=-V(T)\sqrt{1-T_{,\mu}T^{,\mu}}
\end{equation}
 with the potential
$V(T)=\sqrt{A}$ exactly coincides with the Chaplygin gas model
\cite{{fro}, {gori}}.\\
Some experimental data have implied that our universe is not a
perfectly flat universe and recent papers have favoured a universe
with spatial curvature \cite{curve}. As a matter of fact, we want to
remark that although it is believed that our universe is flat, a
contribution to the Friedmann equation from spatial curvature is
still possible if the number of e-foldings is not very large
\cite{miao2}. Cosmic Microwave Background (CMB) anisotropy data
provide the most stringent constraints on cosmic curvature $k$.
Assuming that dark energy is a cosmological constant, the three-year
WMAP data give $\Omega_k=-0.15 \pm 0.11$, and this improves
dramatically to $\Omega_k= -0.005 \pm 0.006$, with the addition of
galaxy survey data from the SDSS \cite{47}. The effect of allowing
non-zero curvature on constraining some dark energy models has been
studied by \cite{curva}. Recently Clarkson et al \cite{clar} have
shown that ignoring $\Omega_k$ induces errors in the reconstructed
dark energy equation of state, $\omega(z)$, that grow very rapidly
with redshift and dominate the $\omega(z)$ error budget at redshifts
$(z \succeq 0.9)$ even if $\Omega_k$ is very small. Due to these
considerations and motivated by the recent work of Chakraborty and
Debnath \cite{c11}, we generalize their work to the non-flat case.
\section{Tachyonic fluid model}
Now we consider the single tachyonic field model, so the action for
the homogeneous tachyon condensate of string theory in a
gravitational background is given by
\begin{equation}\label{E1}
S=\int d^{4}x \sqrt{-g}~\left[\frac{R}{16 \pi G}+\mathcal{L}\right]
\end{equation}
where $R$ and $\mathcal{L}$ are scalar curvature and Lagrangian
density respectively,$\mathcal{L}$ is given by,
\begin{equation}\label{E2}
{\mathcal{L}}=-V(T)\sqrt{1+g^{\mu \nu}
\partial_{\mu}T\partial_{\nu}T},
\end{equation}
where $T$ is tachyon field, and $V(T)$ is the tachyonic potential.
The Fridmann-Robertson-Walker (FRW) metric of universe is as,
\begin{equation}\label{E3}
ds^2 =dt^2-a^2(t) \left[\frac{dr^2}{1-kr^2}+r^2 d\Omega^2\right],
\end{equation}
here $k=1,0,-1$ are corresponds to closed, flat and open universe
respectively.  By using the Einstein's equation we have following
expression
\begin{equation}\label{E4}
\rho_{tot}=\frac{3}{2}\left(\frac{\dot{a}^2}{a^2}+\frac{k}{a^2}\right),
\end{equation}
\begin{equation}\label{E5}
p_{tot}=-\frac{1}{2}\left(2\frac{\ddot{a}}{a}+\frac{\dot{a}^2}{a^2}+\frac{k}{a^2}\right),
\end{equation}
where we have assumed $4\pi G=1$. On the other hand, the
energy-momentum tensor for the tachyonic field is,
\begin{equation}\label{E6}
T_{\mu\nu}= -\frac{2}{\sqrt{-g}}\frac{\delta S}{\delta{g^{\mu
\nu}}}=p_T g_{\mu\nu}+(p_T+\rho_T) u_\mu u_\nu,
\end{equation}
where the velocity $u_\mu$ is
\begin{equation}\label{E7}
u_\mu=- \frac{\partial_{\mu}T}{\sqrt{-g^{\mu \nu} \partial_\mu T
\partial_\nu T}},
\end{equation}
with $u^\nu u_\nu=-1$.\\
By using the energy-momentum tensor we have following expressions
\begin{equation}\label{E8}
R=6\left(\frac{\ddot{a}}{a}+\frac{\dot{a}^2}{a^2}+\frac{k}{a^2}\right),
\end{equation}
\begin{equation}\label{E9-1}
\rho_T=\frac{V(T)}{\sqrt{1-\dot{T}^2}},~~~~~ p_T=-V(T)
\sqrt{1-\dot{T}^2}.
\end{equation}
Thus the equation of states of tachyonic field becomes
\begin{equation}\label{E10}
\omega_T=\frac{p_T}{\rho_T}=\dot{T}^2-1,
\end{equation}
\begin{equation}\label{E9}
p_T \rho_T=-V^2(T),
\end{equation}
Now we consider two fluid model consisting of tachyonic field and
barotropic fluid respectively. The EoS of the barotropic fluid is
given by
\begin{equation}\label{E11}
p_b=\omega_b \rho_b,
\end{equation}
where $p_b$ and $\rho_b$ are the pressure and energy density of
barotropic fluid. Thus, the total energy density and pressure are
respectively given by,
\begin{equation}\label{E12}
\rho_{tot}=\rho_b+\rho_T,
\end{equation}
\begin{equation}\label{E13}
p_{tot}=p_b+p_T,
\end{equation}
 In the next following section we consider two cases, first we
 investigate the case where
two fluid do not interact with each other and second we consider
interacting case.
\section{Non-interacting two fluids model}
In this section we assume that two fluid do not interact with each
other. As we know the general form of conservation equation is
\begin{equation}\label{E14}
\dot{\rho}_{tot}+3\frac{\dot{a}}{a}(\rho_{tot}+p_{tot})=0,
\end{equation}
this equation lead us to write the conservation equation for the
tachyonic and barotropic fluid separately,
\begin{equation}\label{E15}
\dot{\rho}_{b}+3\frac{\dot{a}}{a}(\rho_{b}+p_{b})=0,
\end{equation}
and
\begin{equation}\label{E16}
\dot{\rho}_{T}+3\frac{\dot{a}}{a}(\rho_{T}+p_{T})=0.
\end{equation}
By using the equation (\ref{E15}) one can obtain the energy density
$\rho_b$ as a follow,
\begin{equation}\label{E17}
\rho_b=\rho_0 a^{-3(1+\omega_b)}.
\end{equation}
In order to obtain $T$ and $V(T)$ we first obtain the $\rho_T$ and
$p_T$ in term of $a(T)$,
\begin{equation}\label{E18}
\rho_T=\frac{3}{2}\left(\frac{\dot{a}^2}{a^2}+\frac{k}{a^2}\right)-\rho_0
a^{-3(1+\omega_b)},
\end{equation}
and
\begin{equation}\label{E19}
p_T=-\frac{1}{2}\left(2\frac{\ddot{a}}{a}+\frac{\dot{a}^2}{a^2}+\frac{k}{a^2}\right)-\rho_0
\omega_b a^{-3(1+\omega_b)},
\end{equation}
Here by using the equation (\ref{E10}), (\ref{E18}) and (\ref{E19})
the corresponding field for the tachyon will be as
\begin{equation}\label{E20}
\dot{T}=\sqrt{\frac{\frac{\dot{a}^2}{a^2}+\frac{k}{a^2}-\frac{\ddot{a}}{a}-\rho_0(1+\omega_b)a^{-3(1+\omega_b)}}
{\frac{3}{2}\left(\frac{\dot{a}^2}{a^2}+\frac{k}{a^2}\right)-\rho_0a^{-3(1+\omega_b)}}},
\end{equation}
and from equation (\ref{E9}), the $V(T)$ is given by,
\begin{equation}\label{E21}
V(T)=\sqrt{\left[\frac{1}{2}\left(\frac{\dot{a}^2}{a^2}+\frac{k}{a^2}+2\frac{\ddot{a}}{a}\right)+\rho_0\omega_ba^{-3(1+\omega_b)}\right]
\left[\frac{3}{2}\left(\frac{\dot{a}^2}{a^2}+\frac{k}{a^2}\right)-\rho_0a^{-3(1+\omega_b)}\right]}
\end{equation}
Now we take following ansatz for the scale factor, where increase in
term of time evolution
\begin{equation}\label{E22}
a(t)=\sqrt {b k+c \cosh^2(\beta t)},
\end{equation}
where $b$, $c$ and $\beta$ are constant. By substituting above scal
factor into (\ref{E18}) and (\ref{E19}) the $\rho_T$ and $p_T$ can
be given by following expression,
\begin{eqnarray}\label{E24}
\rho_T=\frac{3}{2}\frac{c \cosh^2(\beta t)\left(c\beta^2 \sinh^2(\beta t)-k\right)+3k^2b}{\left(bk+c \cosh^2(\beta t)\right)^2}\nonumber\\ -\rho_0 \left(bk+c \cosh^2(\beta t)\right)^{-\frac{3}{2}(1+\omega_b)}
\end{eqnarray}
and
\begin{eqnarray}\label{E25}
p_T=-\frac{1}{2}\frac{\cosh^2(\beta t)\left[ 3c^2 \beta^2\cosh^2(\beta t)-c^2 \beta^2+kc+4cbk\beta^2\right]-2cbk\beta^2+k^2 b}{
\left(bk+c \cosh^2(\beta t)\right)^2}\nonumber\\
-\rho_0 \omega_b \left(bk+c \cosh^2(\beta t)\right)^{-\frac{3}{2}(1+\omega_b)}
\end{eqnarray}
By putting the $\rho_T$ and $p_T$ in equation
$\dot{T}=\sqrt{1+\frac{p_T}{\rho_T}}$ and drawing the corresponding
$\dot{T}$ in term of time, we obtain $\dot{T}= \delta ~sech(\eta t)$
where $\delta$ and $\eta$ are given in term of $b$, $c$ and $\beta$.
Then we have,
\begin{equation}\label{E27}
T=\frac{\delta}{\eta} \arctan(\sinh(\eta t)),
\end{equation}
Also, the corresponding potential Eq. (\ref{E21}) will be form of
$A+B e^{-C t^2}$. Now one can use Eq. (\ref{E27}) and reconstruct
potential $V$, in term of tachyon field $T$, as following
\begin{equation}\label{E27-1}
V(T)=A+B ~{\rm e}^{-\frac{C}{\eta^2} \left( {arcsinh} \left( \tan
\left( {\frac {T \eta}{\delta}} \right)  \right) \right) ^{2}},
\end{equation}
By substituting Eqs. (\ref{E24}) and (\ref{E25}) in (\ref{E10}), one
can obtain the equation of state of tachyon field in term of time.
The
 graph of this EoS in
term of time evolution is given by Fig. (1).\\
\begin{tabular*}{2cm}{cc}
\hspace{0.25cm}\includegraphics[scale=0.37]{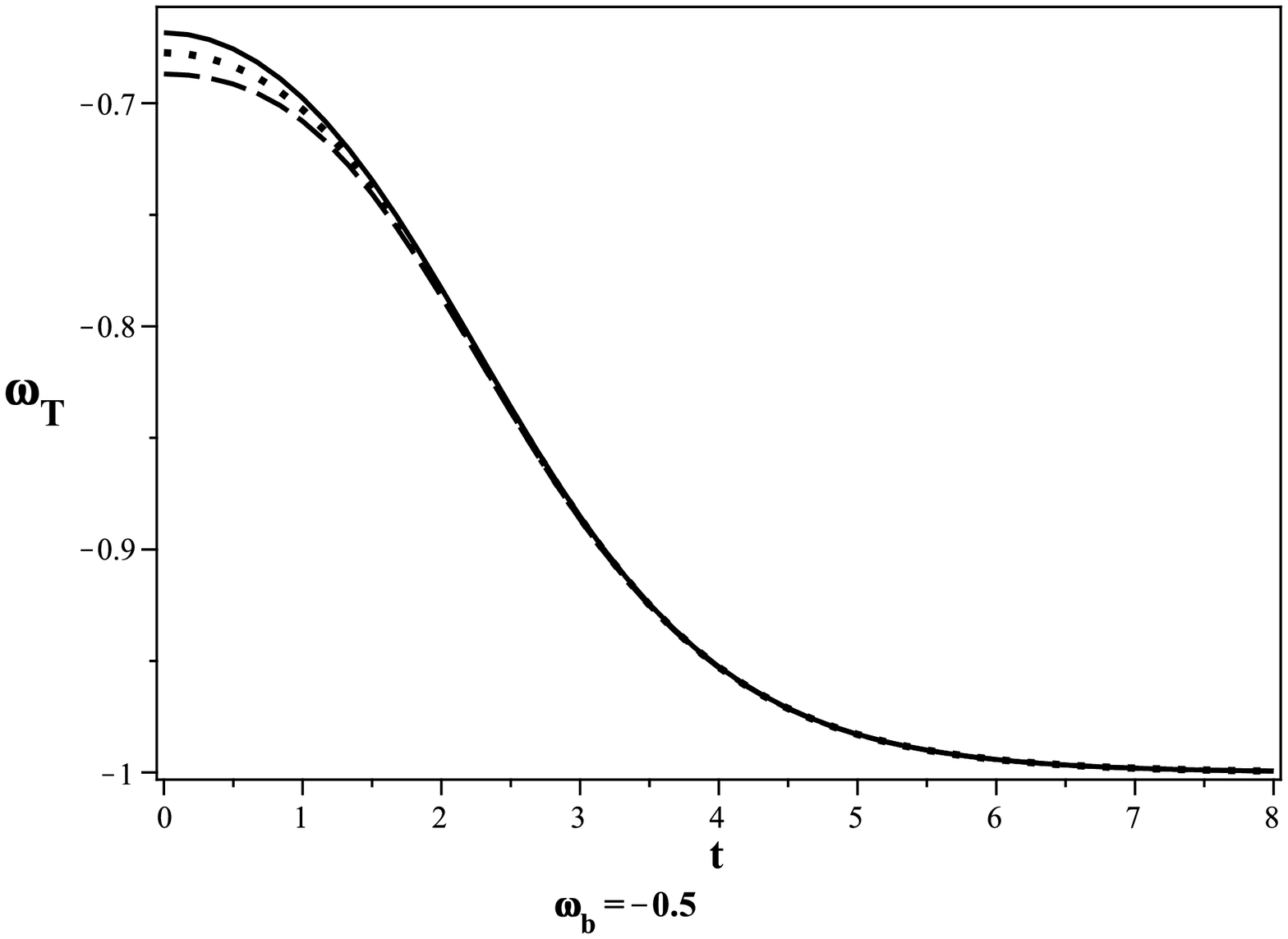}\hspace{0.25cm}\includegraphics[scale=0.355]{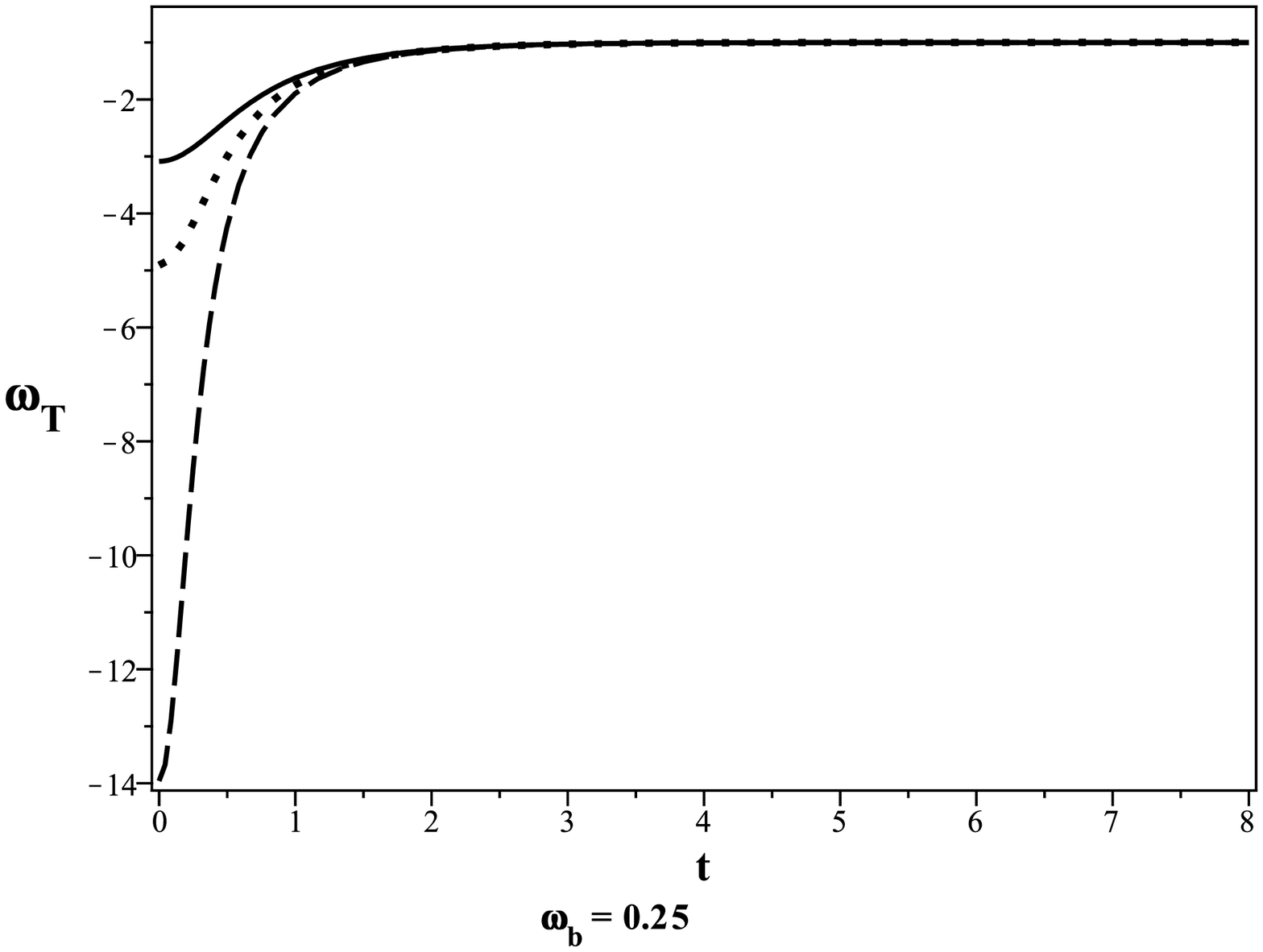}\\
\hspace{2.5cm}Figure 1:\,The EoS is plotted in $\rho_0=-30$, $c=20$, $b=0.5$,
$\beta=0.75$,\\\hspace{2.5cm} and $k=+1, 0, -1$ as
the solid, dot and dash
respectively.\\
\end{tabular*}
\section{Interaction two fluids model}
Now we are going to consider an interaction between the tachyonic
field and the barotropic fluid. Here we need to introduce a
phenomenological coupling function which is a product of the Hubble
parameter and the energy density of the barotropic fluid. In that
case there is an energy flow between the two fluid. so, the equation
of motion corresponding to the tachyonic field and the barotropic
fluid are respectively,
\begin{equation}\label{E28}
\dot{\rho}_{T}+3\frac{\dot{a}}{a}(\rho_{T}+p_{T})=-Q,
\end{equation}
\begin{equation}\label{E29}
\dot{\rho}_{b}+3\frac{\dot{a}}{a}(\rho_{b}+p_{b})=Q,
\end{equation}
where the quantity $Q$ expresses the interaction between the dark
components. The interaction term $Q$ should be positive, i.e. $Q>0$,
which means that there is an energy transfer from the dark energy to
dark matter. The positivity of the interaction term ensures that the
second law of thermodynamics is fulfilled \cite{Pavon:2007gt}.
At this point, it should be stressed that the continuity equations
imply that the interaction term should be a function of a quantity
with units of inverse of time (a first and natural choice can be the
Hubble factor $H$) multiplied  with the energy density. Therefore,
the interaction term could be in any of the following forms: (i)
$Q\propto H\rho_{T}$ \cite{Pavon:2005yx,Pavon:2007gt}, (ii)
$Q\propto H\rho_{b}$ \cite{Amendola:2006dg}, or (iii) $Q\propto
H(\rho_{T}+\rho_{b})$ \cite{Wang:2005ph}. The freedom of choosing
the specific form of the interaction term $Q$  stems from our
incognizance of the origin and nature of dark energy as well as dark
matter. Moreover, a microphysical model describing the interaction
between the dark components of the universe is not available
nowadays. Here we consider $Q=3H \sigma \rho_b$, where $\sigma$ is a
coupling constant. By using the equations (\ref{E28}) and
(\ref{E29}) one can obtain the $\rho_b$, $\rho_T$ and $p_T$ as a
following,
\begin{equation}\label{E30}
\rho_b=\rho_0 a^{-3 (1+\omega_b-\sigma)},
\end{equation}
\begin{equation}\label{E31}
\rho_{T}=\frac{3}{2}\left(\frac{\dot{a}^2}{a^2}+\frac{k}{a^2}\right)-\rho_0
a^{-3(1+\omega_b-\sigma)},
\end{equation}
and
\begin{equation}\label{E32}
p_{T}=-\frac{1}{2}\left(2\frac{\ddot{a}}{a}+\frac{\dot{a}^2}{a^2}+\frac{k}{a^2}\right)-\rho_0
(\omega_b-\sigma) a^{-3(1+\omega_b-\sigma)}.
\end{equation}
Similar to previous case the $\dot{T}$ and $V(T)$ can be obtained by
the following expression,
\begin{equation}\label{E33}
\dot{T}=\sqrt{1-\frac{\frac{1}{2}\left(2\frac{\ddot{a}}{a}+\frac{\dot{a}^2}{a^2}+\frac{k}{a^2}\right)+\rho_0(\omega_b-\sigma)a^{-3(1+\omega_b-\sigma)}}
{\frac{3}{2}\left(\frac{\dot{a}^2}{a^2}+\frac{k}{a^2}\right)-\rho_0a^{-3(1+\omega_b-\sigma)}}},
\end{equation}
and
\begin{equation}\label{E34}
V(T)=\sqrt{\left[\frac{1}{2}\left(2\frac{\ddot{a}}{a}+\frac{\dot{a}^2}{a^2}+\frac{k}{a^2}\right)+\rho_0(\omega_b-\sigma)a^{-3(1+\omega_b-\sigma)}\right]
\left[\frac{3}{2}\left(\frac{\dot{a}^2}{a^2}+\frac{k}{a^2}\right)-\rho_0a^{-3(1+\omega_b-\sigma)}\right]}.
\end{equation}
Putting the value of $a(t)$ from equation (\ref{E22}) in $\rho_T$,
$p_T$, $\dot{T}$ and $V$ we have,
\begin{eqnarray}\label{E35}
\rho_T=\frac{3}{2}\frac{c \cosh^2(\beta t)\left(c\beta^2 \sinh^2(\beta t)-k\right)+3k^2b}{\left(bk+c \cosh^2(\beta t)\right)^2}\nonumber\\ -\rho_0 \left(bk+c \cosh^2(\beta t)\right)^{-\frac{3}{2}(1+\omega_b-\sigma)}
\end{eqnarray}
and
\begin{eqnarray}\label{E36}
p_T=-\frac{1}{2}\frac{\cosh^2(\beta t)\left[ 3c^2 \beta^2\cosh^2(\beta t)-c^2 \beta^2+kc+4cbk\beta^2\right]-2cbk\beta^2+k^2 b}{
\left(bk+c \cosh^2(\beta t)\right)^2}\nonumber\\
-\rho_0 (\omega_b-\sigma) \left(bk+c \cosh^2(\beta t)\right)^{-\frac{3}{2}(1+\omega_b-\sigma)}
\end{eqnarray}
By substituting the $\rho_T$ and $p_T$ in equation
$\dot{T}=\sqrt{1+\frac{p_T}{\rho_T}}$ and drawing the corresponding
$\dot{T}$ in term of time we have graphs of $T$ and $V$ as the same
results with non-interaction case.\\ By substituting Eqs.
(\ref{E35}) and (\ref{E36}) in (\ref{E10}), graphs of the EoS are
given in term
of time evolution in Fig. (2).\\
\begin{tabular*}{2cm}{cc}
\hspace{0.25cm}\includegraphics[scale=0.37]{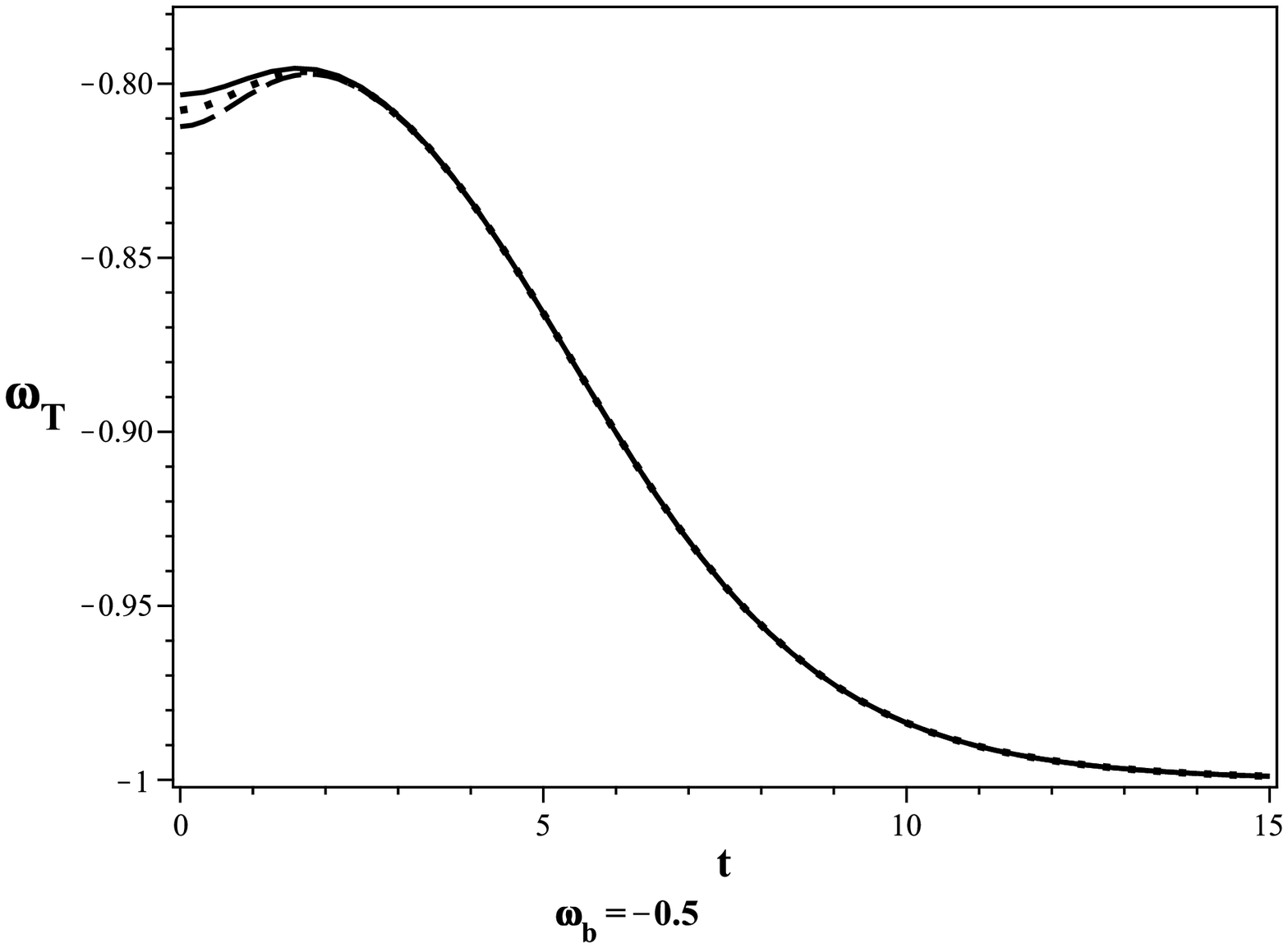}\hspace{0.25cm}\includegraphics[scale=0.355]{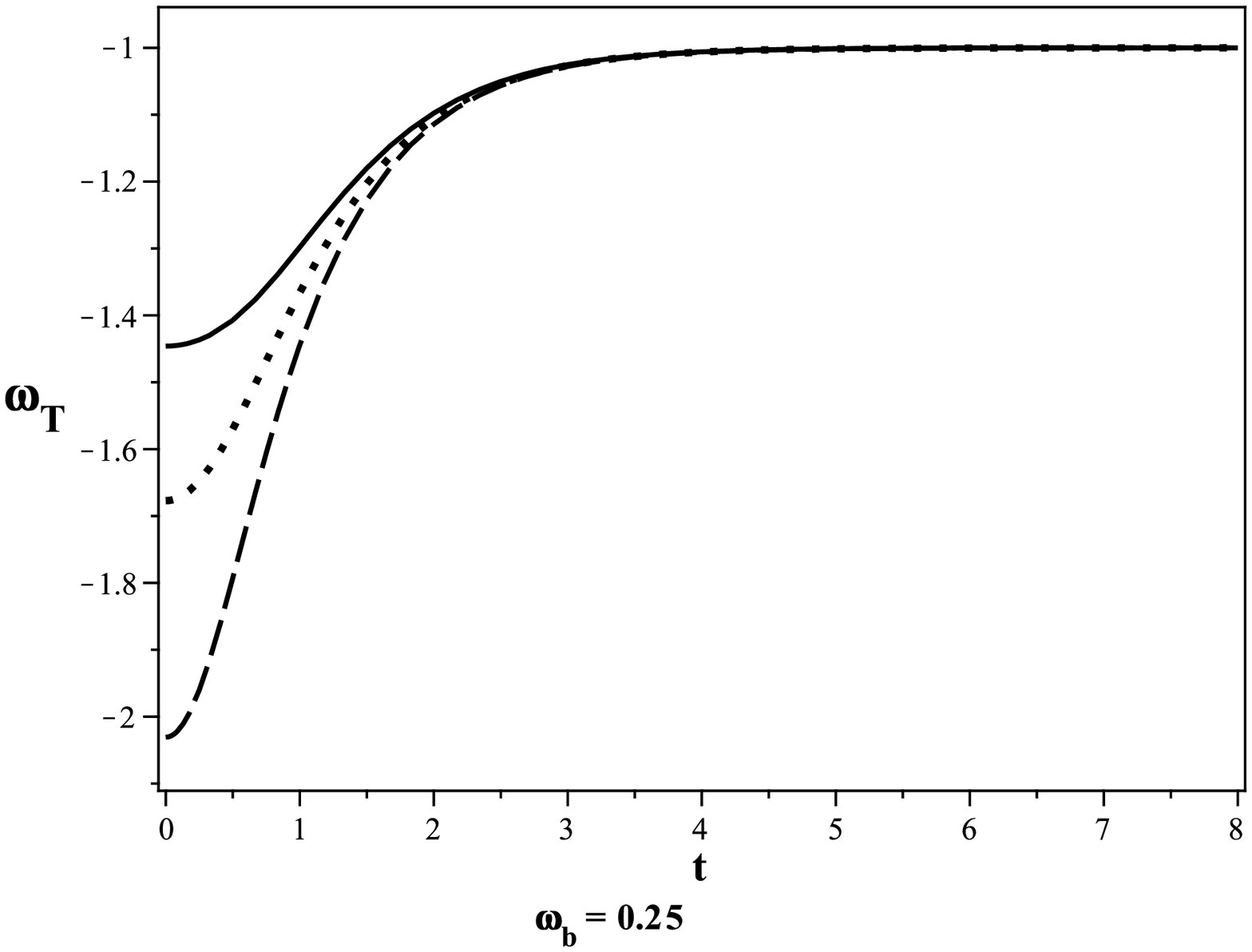}\\
\hspace{2cm}Figure 2:\,The EoS is plotted in $\rho_0=-30$, $c=20$, $b=0.5$,
$\beta=0.75$,\\\hspace{2cm} $\sigma=0.25$  and $k=+1,
0, -1$ as the solid, dot and dash
respectively.\\
\end{tabular*}

\section{Conclusion}
Within the different candidates to play the role of the dark energy,
tachyon, has emerged as a possible source of dark energy for a
particular class of potentials \cite{tac}. In the present paper we
have studied the tachyonic field model in non-flat universe. At
first we have considered the non-interacting case, where the tachyon
field and barotropic fluid separately satisfy the conservation
equation. We have obtained the evaluation of scale factor, energy
density, pressure, tachyon field and potential of tachyon field in
term of cosmic time. After that we have reconstruct the potential of
tachyon in term of tachyon field. The evaluation of EoS, and the
late time behavior of this equation is given by fig 1.\\
Studying the interaction between the dark energy and ordinary matter
will open a possibility of detecting the dark energy. It should be
pointed out that evidence was recently provided by the Abell Cluster
A586 in support of the interaction between dark energy and dark
matter \cite{Bertolami:2007zm}. However, despite the fact that
numerous works have been performed till now, there are no strong
observational bounds on the strength of this interaction
\cite{Feng:2007wn}. This weakness to set stringent (observational or
theoretical) constraints on the strength of the coupling between
dark energy and dark matter stems from our unawareness of the nature
and origin of dark components of the Universe. It is therefore more
than obvious that further work is needed to this direction. Due to
this we have extended the our consideration to the interacting case
in a separate section of this paper.

\section{Acknowledgement} The authors would like to thank
an anonymous referee for crucial remarks and advices.

\end{document}